\documentclass[10pt]{article}
\usepackage{multirow}
\usepackage{calc}
\usepackage{epsfig,amsfonts,amsthm,latexsym,amsmath}

\newcommand{\mat}[1]{\mbox{\boldmath{$#1$}}}

\usepackage{subfigure} 

\usepackage{makeidx}     
\usepackage{graphicx}    
\usepackage{multicol}
\usepackage{amssymb}
\usepackage{latexsym}
\usepackage{amsmath}
\usepackage{amsfonts}
\usepackage{amssymb}
\usepackage{amsopn}
\usepackage{amsthm}
\usepackage{multirow}
\usepackage{scalefnt}
\usepackage[latin1]{inputenc}

\usepackage[usenames]{color} 

\theoremstyle{plain}

\theoremstyle{remark}

\title{\normalsize\bf
ON MODELING OF VARIABILITY IN MIXTURE EXPERIMENTS WITH NOISE VARIABLES
} 

\author{
Edmilson Rodrigues Pinto$^{1*}$, \  \ Leandro Alves Pereira$^{1}$  \\ \ and \ Aurélia A. de Araújo Rodrigues$^{1}$
}

\begin{document}

\date{}

\maketitle

\vspace{-20pt}
\begin{center}
{\footnotesize 
*Corresponding author\\
$^1$Faculty of Mathematics, Federal University of Uberl\^andia - Brazil.\\

E-mails: edmilson@famat.ufu.br / leandro@famat.ufu.br/ aurelia@famat.ufu.br  
}\end{center}

\bigskip
\noindent
{\small{\bf ABSTRACT.}
In mixture experiments with noise variables or process variables that can not be controlled, investigate and try to control the variability of the response variable is very important for quality improvement in industrial processes. Thus, modeling the variability in mixture experiments with noise variables becomes necessary and has been considered in literature with approaches that require the choice of a quadratic loss function or by using the delta method. In this paper, we make use of the delta method and also propose an alternative approach, which is based on the Joint Modeling of Mean and Dispersion (JMMD). We consider a mixture experiment involving noise variables and we use the techniques of JMMD and of the delta method to get models for both mean and variance of the response variable. Following the Taguchi's ideas about robust parameter design we build and solve an optimization problem for minimizing the variance while holding the mean on the target. At the end we provide a discussion about the two methodologies considered.
}


\baselineskip=\normalbaselineskip


\section{Introduction}
 \label{introd} 

Experiments with mixtures involve the mixing or blending of two or more ingredients to form an end product. For this type of experiment it is of interest to determine the proportions of the mixture components which lead to desirable results with respect to some quality characteristic of interest.

In general, the quality of the end product is a function of the proportions of the ingredients and of other factors that do not form any portion of the mixture, as heat or time; these factors are called process variables or process conditions and can not always be controlled. In other words, in some mixture experiments, the response depends not only on the proportion of the mixture components present in the mixture but also on the processing conditions that are, in general, designated as process variables. Process variables are factors in an experiment that do not form any portion of the mixture but whose levels when changed could affect the blending properties of the ingredients. In the literature on mixture experiments with process variables, the goal is determining the proportions of the mixture components along with situations of process conditions, so that the response becomes as close as possible to a target value.

Experiments with mixture and process variables are well covered in \cite{Cornell}, but modeling of variance is not considered. The modeling of variance in mixture experiments with noise variables has been considered in \cite{SteinerHamada}, who proposed a combined mixture-process-noise variable model, built and solved an optimization problem to minimize a quadratic loss function, taking into account both mean and variance of response. Another approach to modeling the variance in mixture experiments is due to \cite{Goldfarb} using the delta method, which employs a Taylor series approximation of the regression model at a vector of process variables.

In this paper, besides the delta method, we also consider the joint modeling of mean and dispersion (JMMD) as an alternative approach to modeling the variance in mixture experiments with noise variables. The JMMD was introduced by \cite{NelderLee} as an alternative to Taguchi's methods in quality-improvement experiment and provides a methodology to find and check the fit of the models found with a solid statistical basis. Further examples of applications of the JMMD can be found in \cite{LeeNelderI} and \cite{LeeNelderII}. A a comprehensive review of the joint modeling of mean and dispersion proposed by \cite{NelderLee} is given by \cite{PintoPonce}.

We consider a mixture experiment involving noise variables and we use the approaches of JMMD and the delta method to get models for both mean and variance of the response variable. Following the Taguchi's ideas about robust parameter design, see \cite{Taguchi} we build and solve an optimization problem for minimizing the variance while holding the mean on the target. At the end we present some considerations about both methods used.

The paper is organized as follows. In Section 2, we introduce mixture experiments and present some models used for mixtures with process variables. In Section 3, we describe briefly the joint modeling of mean and dispersion and discuss the principal points of the theory. In Section 4, we make a resume about the delta method. In Section 5, we apply the JMMD and the delta method to get models for both mean and variance in an example of mixture experiment involving noise variables. Additionally we build and solve an optimization problem for minimizing the variance while holding the mean on the target. Finally, in Section 6, we provide a discussion about the two methodologies considered.

\section{Mixture experiments} 
\label{MixtureExperiments}

A mixture experiment involves mixing proportions of two or more components to make different compositions of an end product. Mixture component proportions $x_i$ are subject to the constraints

\begin{equation} \label{eq1}
0\le x_i \le 1 \quad \quad i=1,2,\dots,a \quad \quad \textrm{and} \quad \quad \sum_{i=1}^{a} x_{i}=1
\end{equation}

\noindent where $a$ is the number of components involved in the mixture experiment. Consequently, the design space is a $(a-1)$-dimensional simplex or part of a simplex if there are further conditions on the proportions such as $l_i \le x_i \le u_i$ for $i=1,2, \dots, a-1$, with the proportion $x_a$ taking values which make up the mixture.

If, in addition to the $a$ mixture components $\mat{x}^t=(x_1,\dots,x_{a})$, there are $b$ process variables $\mat{z}^t=(z_1,\dots,z_b)$; we can consider typically additive models like $\eta(\mat{x},\mat{z})=\zeta(\mat{x})+\vartheta(\mat{z})$ or complete cross product models of the type $\eta(\mat{x},\mat{z})=\zeta(\mat{x})\vartheta(\mat{z})$ or combinations of these such as $\eta(\mat{x},\mat{z})=\zeta(\mat{x})+ \nu(\mat{x},\mat{z})$, where $\zeta(\mat{x})$ represents the mixture model, $\vartheta(\mat{z})$ represents the process variable model and $\nu(\mat{x},\mat{z})$ comprises products of terms in $\zeta(\mat{x})$ and $\vartheta(\mat{z})$.
For three mixture components and two process variables, with $\zeta(\mat{x})= \sum_{i=1}^{3}\beta_i x_i + \sum_{i=1}^{2}\sum_{j>i}^{3}\beta_{ij} x_i x_j + \beta_{123}x_1 x_2 x_3 + \sum_{i=1}^{2} \sum_{j>i}^{3} \gamma_{ij}x_i x_j (x_i - x_j)$ and $\vartheta(\mat{z})=\alpha_{0}+\sum_{i=1}^{2}\alpha_{i}z_{i}+\sum_{i=1}^{2}\alpha_{ii}z_{i}^{2}+ \alpha_{12}z_{1}z_{2}$, the combined multiplicative model, which includes the Scheff\'e cubic model for the mixture and the reduced quadratic model for the process variables, is given by

\begin{eqnarray} \label{eq2}
\eta(\mat{x},\mat{z})& = &\sum_{i=1}^{3}\beta^{0}_i x_i + \sum_{i=1}^{2}\sum_{j>i}^{3}\beta^{0}_{ij} x_i x_j + \beta^{0}_{123}x_1 x_2 x_3 + \sum_{i=1}^{2} \sum_{j>i}^{3} \gamma^{0}_{ij}x_i x_j (x_i -
 \nonumber \\
                     &   & x_j) + \sum_{l=1}^{2} \left[\sum_{i=1}^{3}\beta^{l}_i x_i + \sum_{i=1}^{2}\sum_{j>i}^{3}\beta^{l}_{ij} x_i x_j + \beta^{l}_{123}x_1 x_2 x_3 + 
  \right.  \nonumber \\
                     &   & \left. \sum_{i=1}^{2} \sum_{j>i}^{3} \gamma^{l}_{ij}x_i x_j (x_i - x_j) \right]z_{l} + \sum_{l=1}^{2}      \left[\sum_{i=1}^{3}\beta^{ll}_i x_i + \sum_{i=1}^{2}\sum_{j>i}^{3}\beta^{ll}_{ij} x_i x_j +  
  \right. \nonumber \\
                     &   & \left. \beta^{ll}_{123}x_1 x_2 x_3 + \sum_{i=1}^{2} \sum_{j>i}^{3} \gamma^{ll}_{ij}x_i x_j (x_i - x_j) \right]z^{2}_{l} + \left[\sum_{i=1}^{3}\beta^{12}_i x_i + 
\right. \nonumber \\ 
                     &   & \left. \sum_{i=1}^{2}\sum_{j>i}^{3}\beta^{12}_{ij} x_i x_j + \beta^{12}_{123}x_1 x_2 x_3 + \sum_{i=1}^{2} \sum_{j>i}^{3} \gamma^{12}_{ij}x_i x_j (x_i - x_j) \right]z_{1}z_{2} \qquad                                      
\end{eqnarray}

In general, the methodology used to construct mixture designs involving process variables is combination of two designs, one being a mixture design for the mixture components and the other being factorial or fractional factorial design for the process variables. For more details about mixture experiments with process variables see \cite{Cornell}.

\section{Joint modeling of mean and dispersion} 
\label{JMMD}

According to \cite{NelderLee}, the method of joint modeling of mean and dispersion consists of finding joint models for the mean and dispersion. In their approach, using the extended quasi-likelihood, two interlinked generalized linear models (GLM) are needed, one for the mean and the other for the dispersion. For the dispersion model is used as response variable the deviance component

\begin {equation} \label{eq3}
d_{i} = 2 \int_{\mu_{i}}^{y_{i}}\frac{(y_i-\ell)}{V(\ell)}d\ell,   
\end{equation}

\noindent for each observation $y_i$, where $\mu_i$ represents the mean for the $i$th observation and $V(.)$ is the variance function for the GLM, see \cite{McCullaghNelder}, p. 360. 

Let $\mat{x}^t=(x_1,\ldots,x_a)$ be a vector with design factors and $\mat{z}^t=(z_1,\ldots,z_b)$ be a vector with noise factors. Suppose that $\mat{t}^t=(t_1,\ldots,t_s)$ and $\mat{u}^t=(u_1,\ldots,u_r)$ are the independent variables that affect the mean and the dispersion models respectively. The vectors $\mat{t}$ and $\mat{u}$ may contain factors occurring in $\mat{x}$, in $\mat{z}$ or in both. The independent variables for the dispersion model are commonly, but not necessarily, a subset of the independent variables for the mean model. 

Consider $Y_1, \ldots,Y_n$ $n$ independent random variables with the same probability distribution, whose values, given by $y_1,\ldots,y_n$, are the results of an experimental arrangement. Spite of the distribution of $Y_i$ is unknown, it is assumed that $E(Y_i|\mat{Z}_i)=\mu_i$ and $Var(Y_i| \mat{Z}_i)=\phi_iV(\mu_i)$, where $ \mat{Z}_i$ is a vector with the fixed values for the random noise variables, $\phi_i$ is the dispersion parameter and $V(.)$ is the variance function. Let $\varphi$ be a link function for the mean model, i.e., $\eta_i=\varphi(\mu_i)=\mat{f}^t(\mat{t}_i)\mat{\beta}$ with $\mat{f}^t(\mat{t}_i)=(f_1(\mat{t}_i),\ldots,f_p(\mat{t}_i))$ where $f_j(\mat{t}_i)$, for $j=1,\ldots,p$, is a known function of $\mat{t}_i$ and $\mat{\beta}$ is a $p\times1$ vector of unknown parameters. Following \cite{LeeNelderI}, for the dispersion model we are assuming a gamma model with a log link function, i.e., $\tau_i=\psi(\phi_i)=\ln(\phi_i)=\mat{g}^t(\mat{u}_i)\mat{\gamma}$, with $\mat{g}^t(\mat{u}_i)=(g_1(\mat{u}_i),\ldots,g_q(\mat{u}_i))$, where $g_j(\mat{u}_i)$, for $j=1,\ldots,q$, is a known function of $\mat{u}_i$ and $\mat{\gamma}$ is a $q\times1$ vector of unknown parameters. Thus, on the dispersion model we are considering $E(d_i)=\phi_i$ and $Var(d_i)=2\phi_{i}^{2}$ (see \cite{McCullaghNelder}, p. 361). Note that, in general, the factors occurring in $\mat{f}(\mat{t})$ and $\mat{g}(\mat{u})$ can be linear effects, quadratic effects or interactions between the factors occurring in $\mat{x}$ and in $\mat{z}$. A term occurring in the mean linear predictor only can thus be used to get the mean close to target, while a term in the dispersion linear predictor, whether or not it occurs also in the mean, can be used to reduce the dispersion. We also define $\mat{T}$ and $\mat{U}$ the experimental matrices for the mean and dispersion models respectively, with $\mat{T}=[\mat{f}(\mat{t}_1),\ldots,\mat{f}(\mat{t}_n)]^t$ and $\mat{U}=[\mat{g}(\mat{u}_1),\ldots,\mat{g}(\mat{u}_n)]^t$, where $n$ is the number of observations.

The fitting for the JMMD uses as an optimizing criterion the idea of extended quasi-likelihood, introduced by \cite{NelderPregibon}, see \cite{McCullaghNelder}, p. 349. In this work we use the adjusted extended quasi-likelihood, introduced by \cite{LeeNelderI}. The adjusted extended quasi-likelihood is given, in our notation, by

\begin{equation} \label{eq4}
-2Q_{A}^{+}=\sum_{i=1}^{n}\left( \frac{d_{i}^{*}}{\phi_{i}} + \ln \{ 2\pi\phi_{i}V(y_{i})\}\right)
\end{equation}

\noindent where $d_{i}^{*}=\frac{d_{i}}{1-h_{i}}$ is the standardized deviance component and $h_{i}$ is the $i$th element of the diagonal of the matrix $\mat{H}=\mat{W}^{\frac{1}{2}}\mat{T}(\mat{T}^{t}\mat{WT})^{-1}\mat{T}\mat{W}^{\frac{1}{2}}$, being $\mat{W}$, the weight matrix for the GLM, a diagonal matrix with elements given by $w_{i}=\left(\frac{\partial\mu_{i}}{\partial\eta_{i}}\right)^{2}\frac{1}{\phi_i V(\mu_{i})}$. The Table \ref{Tab1} gives a resume of the joint modeling of mean and dispersion. From Table \ref{Tab1}, we can observe that the standardized deviance component from the model for the mean becomes the response for the dispersion model, and the inverse of fitted values for the dispersion model provides the prior weights for the mean model.

\begin {table}[htb]
\caption{Summary of the JMMD}
\begin{tabular}{p{3.5cm} p{3.5cm} p{5cm}}
\hline
Component               & Mean model                                           & Dispersion model$^{\dag}$                        \\ \hline
Response variable       &  $y_i$                                               &  $d_{i}^{*}$                                     \\
Mean                    &  $\mu_i$                                             &  $\phi_i$                                        \\ 
Variance                &  $\phi_i V(\mu_i)$                                   &  $2\phi_{i}^{2}$                                 \\
Link function           &  $\eta_i=\varphi(\mu_i)$                             &  $\xi_i=\ln(\phi_i)$                             \\
Linear predictor        &  $\eta_i=\mat{f}^t(\mat{t}_i)\mat{\beta}$            &  $\xi_i=\mat{g}^t(\mat{u}_i)\mat{\gamma}$        \\
Deviance component      &  $d_i=2 \int_{\mu_i}^{y_i}\frac{y_i-e}{V(e)}de $     &  
$d_{d_i}=2\left\{-\ln\left(\frac{d_{i}^{*}}{\phi_i}\right) + \frac{(d_{i}^{*}-\phi_i)}{\phi_i}\right\}$                           \\
Prior weight            &  $\frac{1}{\phi_i}$                                  &  $(1-h_i)/2$                                     \\ \hline
\multicolumn{3}{l}{$^{\dag}$\scriptsize{For the dispersion model we are assuming a gamma model with logarithmic link function}}
\end{tabular}
\label{Tab1}
\end{table}

The algorithm for the JMMD is an extension of the standard GLM algorithm, in which the model for the mean is fitted assuming that the fitted values for the dispersion are known and that the model for dispersion is fitted using the fitted values for the mean. The fitting alternates between the mean and dispersion models until convergence is achieved.

After complete convergence, that is, if the equation (\ref{eq16}) is satisfied for a small $\epsilon$, and after model checking the final joint model is found. The iterative process and the algorithm for the JMMD, adapted from \cite{PintoPonce}, are shown in Appendices A and B.

Note that the models for mean and variance, conditional to random noise variables, are obtained as $E(Y|\mat{Z})=\mu=\varphi^{-1}\left[\eta(\mat{x},\mat{Z})\right]$ and $Var(Y|\mat{Z})=\phi V(\mu)=\exp\left\{\xi(\mat{x},\mat{Z})\right\}V(\mu)$. We can find $E(Y)$ and $Var(Y)$ by using the expressions $E(Y)=E(E(Y|\mat{Z}))$ and $Var(Y)= Var(E(Y|\mat{Z}))+ E(Var(Y|\mat{Z}))$. It is assumed that the  distribution of $\mat{Z}$ is known or can be estimated.

\section{Delta method}
\label{delta-method}

The delta method is a well-known technique for finding approximations, based on Taylor series expansions, to the mean and variance of functions of random variables. In this paper, the delta method will be applied, considering the second-order expansion of Taylor series around the mean of noise variables. Let $Z_1,\ldots,Z_b$ be random variables with means $\mu_1,\ldots,\mu_b$ and define $\mat{Z}^t=(Z_1,\ldots,Z_b)$ and $\mat{\mu}^t=(\mu_1,\ldots,\mu_b)$. For our problem of mixture experiments, we also consider $\mat{x}$ as the vector of mixture components. Suppose there is a twice differentiable function $\eta(\mat{x},\mat{Z})$ for which we want an approximate estimate of mean and variance. Define $\eta'\left(\mat{x},\mat{\mu}\right)=\left.\displaystyle\frac{\partial}{\partial \mat{z}} \eta\left(\mat{x},\mat{z}\right)\right|_{\mat{z}=\mat{\mu}}$ and $\mat{\cal{H}}=\left.\displaystyle\frac{\partial^{2}}{\partial \mat{z}\partial \mat{z}^t} \eta \left(\mat{x},\mat{z}\right)\right|_{\mat{z}=\mat{\mu}}$. The second-order Taylor series expansion of $\eta$ about $\mat{\mu}$ is

\begin{equation} \label{eq5}
\eta(\mat{x},\mat{z})=\eta(\mat{x},\mat{\mu})+(\mat{z}-\mat{\mu})^t \eta'(\mat{x},\mat{\mu})+(\mat{z}-\mat{\mu})^t\mat{\cal{H}}(\mat{z}-\mat{\mu})+ \cal{R},
\end{equation}

where $\cal{R}$ is the remainder of the approximation and will be ignored. Thus, taking $Y=\eta(\mat{x},\mat{Z})$ we can obtain

\begin{equation} \label{eq6}
E(Y)=E\left[\eta(\mat{x},\mat{Z})\right]=\eta(\mat{x},\mat{\mu})+E\left[(\mat{Z}-\mat{\mu})^t\mat{\cal{H}}(\mat{Z}-\mat{\mu})\right]
\end{equation} and

\begin{equation} \label{eq7}
Var(Y)=Var\left[\eta(\mat{x},\mat{Z})\right]=Var\left[(\mat{Z}-\mat{\mu})^t \eta'(\mat{x},\mat{\mu})+(\mat{Z}-\mat{\mu})^t\mat{\cal{H}}(\mat{Z}-\mat{\mu})\right].
\end{equation}

For a comprehensive treatment about the delta method see \cite{CasellaBerger}, p. 240.  

\section{Application to a bread-making problem}
\label{bread-making-problem}

The bread-making problem, originally presented by \cite{FaergestadNaes}, according to \cite{NaesFaergestadCornell}, consisted of an experiment with three ingredients of mixture and two noise variables, and had as objective to investigate and to value the final quality of flour, composed by different mixtures of wheat flour, for production of bread. \cite{FaergestadNaes}, according to \cite{NaesFaergestadCornell}, considered three types of wheat flour: two Norwegian, Tjalve $(x_{1})$ and Folke $(x_{2})$ and one American, Hard Red Spring $(x_{3})$; that were considered as control variables, and two types of noise variables: mixing time $(z_{1})$ and the proofing (resting) times of the dough $(z_{2})$, considered as noise variables. The response variable was considered as the loaf volume after baking with target value of 530 ml. The flour blends were considered to be mixing ingredients with $x_{1} + x_{2} +x_{3} = 1$ and with constraints $0.25\le x_{1} \le 1.0$; $0 \le x_{2} \le 0.75$ and $0 \le x_{3} \le 0.75$, where $x_{1}$, $x_{2}$ and $x_{3}$ are the proportions of Tjalve, Folke and Hard Red Spring flour, respectively. For the noise variables, it was considered three situations for the mixing time: 5, 15 and 25 minutes and also three situations for proofing time: 35, 47.5, and 60 minutes. 

\begin{figure}[!htb]
\centering
\includegraphics[scale=0.8]{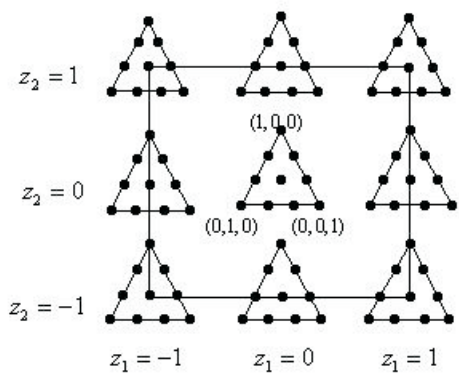}
\caption{\scriptsize The complete design in two noise variables, $z_{1}$ and $z_{2}$, and a simplex-lattice design in three mixture components $x_{1}$, $x_{2}$ and $x_{3}$.}
\label{fig1}
\end{figure}

A full $3^{2}$ factorial design was used for the noise variables and the 10 runs corresponding to a simplex lattice design were replicated at each of the nine combinations of the mixing and proofing times, so that the complete design involved 90 experimental runs as shown in Figure \ref{fig1}. We consider the noise variables coded as $z_{1}=(\textrm{mixing time}-15)/10$ and $z_{2}=(\textrm{proofing time}-47.5)/12.5$.

The volumes recorded for the 10 flour types and the 9 combinations of the noise variables are reproduced in Table \ref{Tab2}, using the run numbers, from 1 to 10, as identified in Figure \ref{fig1}. Additional details of the way in which the experiment was conducted are given by \cite{NaesFaergestadCornell}, and further description of the practical aspects of the study is provided by \cite{FaergestadNaes} according to \cite{NaesFaergestadCornell}.

\begin {table}[htb]
\caption{Loaf volume for the 10 flour types and the 9 combinations of the noise variables}

\label{Tab2}
\tiny
\centering
\begin{tabular}{p{0.35cm} p{0.35cm} p{0.35cm} p{0.3cm} p{0.3cm} p{0.35cm} p{0.65cm} p{0.6cm} p{0.6cm} p{0.6cm} p{0.6cm} p{0.6cm} p{0.6cm} p{0.6cm} p{0.6cm}}
\hline                          
& & & & & &\multicolumn{9}{c}{\scriptsize{Noise Factors}} \\
\cline{7-15}
&\multicolumn{3}{c}{\scriptsize{Design factors}}& &$z_{1}$ & -1 &  0 &  1 &-1 & 0 & 1 & -1& 0 & 1 \\
\cline{2-4}  
$\textrm{n}^{\textrm{o}}$& $x_{1}$& $x_{2}$& $x_{3}$& & $z_{2}$ & -1 & -1 & -1 & 0 & 0 & 0 & 1 & 1 & 1 \\
\hline
1 & 0.25 & 0.75 & 0.00 & & & 378.89 & 396.67 &392.22&445.56&452.22&487.78&457.22&500.56&472.78\\  
2 & 0.50 & 0.50 & 0.00 & & & 388.89 & 423.33 &416.11&460.00&488.89&475.78&472.78&478.00&506.11\\  
3 & 0.75 & 0.25 & 0.00 & & & 426.11 & 483.33 &389.44&474.44&514.44&462.78&506.67&591.67&522.22\\  
4 & 1.00 & 0.00 & 0.00 & & & 386.11 & 459.11 &423.33&458.33&506.11&514.44&545.56&522.22&551.11\\  
5 & 0.25 & 0.50 & 0.25 & & & 417.78 & 437.22 &444.56&484.44&490.00&495.00&497.78&531.11&577.78\\  
6 & 0.50 & 0.25 & 0.25 & & & 389.44 & 447.22 &415.00&490.89&528.89&507.78&517.78&567.22&538.33\\  
7 & 0.75 & 0.00 & 0.25 & & & 448.33 & 459.44 &455.56&436.00&535.00&552.22&507.44&578.89&590.00\\  
8 & 0.25 & 0.25 & 0.50 & & & 413.89 & 485.56 &462.22&483.89&529.44&540.00&565.00&598.89&580.56\\  
9 & 0.50 & 0.00 & 0.50 & & & 415.56 & 514.44 &437.78&493.89&583.33&578.89&524.44&694.44&640.00\\  
10& 0.25 & 0.00 & 0.75 & & & 432.78 & 498.33 &517.22&474.44&568.33&579.44&541.11&638.89&638.89\\  
\hline
\end{tabular}
\end{table}

Naes et al. \cite{NaesFaergestadCornell} carried out a study of the use of robust design methodology to the bread-making problem to investigate the underlying relationships between the response variable loaf volume and the mixture and noise variables, comparing three techniques for analysing the loaf volume, i.e., the mean square error, the analysis of variance and the regression approach, where all factors, the three mixtures components and the two noise variables, are modeled simultaneously. In the analysis carried out by them, the full crossed model (\ref{eq2}) for three mixture ingredients and two noise variables was taken as the starting model, however a detailed argument was presented for reducing the number of parameters by removing some of the second and third-order mixture terms before performing the cross. So the initial reduced model had 28 terms. After Backwards elimination using regression methods, the final model with 18 terms was obtained. Naes et al. \cite{NaesFaergestadCornell} obtain their results considering a homoscedastic Gaussian model for the response variable.

In this paper, we reexamine the bread-making problem considering the possibility of obtaining, in addition to the mean model, a model for variance. In our analysis, we will consider the Taguchi's ideas on Robust Parameter Design (RPD). According to \cite{Montgomery}, RPD is an approach to product realization activities that emphasizes choosing the levels of controllable factors (or parameters) in a process or product to achieve two objectives: i) to ensure that the mean of the output response is at a desired level or target and ii) to ensure that the variability around this target value is as small as possible. In a problem of robust parameter design we must obtain models for both mean and variance of the process or product. Thus, we can minimize the variability by finding the optimum settings of factors that affect the variance model and then adjusting the mean to its target value by finding appropriate settings of factors that affect the mean model. For our purpose, we will use two distinct methodologies to get the models for mean and variance: the joint modeling of mean and dispersion and the delta method. 

\subsection{Modeling of mean and variance via JMMD}

For we apply the JMMD we need to choose a variance function and a link function for the mean model, besides the independent variables for the models of the mean and dispersion. Naes et al. \cite{NaesFaergestadCornell} have considered a Gaussian model in their analysis; thus, initially we consider the model for the mean like Gaussian with identity link function and variance function $V(\mu)=1$. For the dispersion model, as mentioned before in Section \ref{JMMD}, it is assumed a gamma model with logarithmic link function. These assumptions will be checked in our analysis.

\subsubsection{Model building strategy} \vspace{0.3cm}
\label{Model_building_strategy}

We start our analysis with the same initial reduced model, involving 28 terms, considered by  \cite{NaesFaergestadCornell} and given in equation (\ref{eq8}). 

\begin{eqnarray} \label{eq8}
\eta(\mat{x},\mat{z})  & = &\beta^{0}_{1}x_{1} + \beta^{0}_{2}x_{2} + \beta^{0}_{3}x_{3} + \beta^{0}_{12}x_{1}x_{2} + \beta^{0}_{13}x_{1}x_{3} + 		                    \gamma^{0}_{12}x_{1}x_{2}(x_{1}-x_{2})+\nonumber \\
          &   &\gamma^{0}_{13} x_{1}x_{3}(x_{1}-x_{3})+ \{ \beta^{1}_{1}x_{1} + \beta^{1}_{2}x_{2} + \beta^{1}_{3}x_{3} + 
                \beta^{1}_{12}x_{1}x_{2} + \beta^{1}_{13}x_{1}x_{3} + \nonumber \\
          &   &\gamma^{1}_{12}x_{1}x_{2}(x_{1}-x_{2}) + \gamma^{1}_{13}x_{1}x_{3}(x_{1}-x_{3}) \}z_{1} + \{ \beta^{2}_{1}x_{1} +
                \beta^{2}_{2}x_{2} + \nonumber \\
          &   &\beta^{2}_{3}x_{3} + \beta^{2}_{12}x_{1}x_{2} +  \beta^{2}_{13}x_{1}x_{3} + \gamma^{2}_{12}x_{1}x_{2}(x_{1}-x_{2}) + 
                \nonumber \\
          &   &\gamma^{2}_{13}x_{1}x_{3}(x_{1}-x_{3}) \}z_{2}+ \{ \beta^{11}_{1}x_{1} + \beta^{11}_{2}x_{2} + \beta^{11}_{3}x_{3} +  
          	  \beta^{11}_{12}x_{1}x_{2} +   \nonumber \\
          &   &\beta^{11}_{13}x_{1}x_{3} +\gamma^{11}_{12}x_{1}x_{2}(x_{1}-x_{2}) + \gamma^{11}_{13}x_{1}x_{3}(x_{1}-x_{3}) \}z^{2}_{1} 
\end{eqnarray}

We consider a linear regression model for the mean fitted by Ordinary Least Squares (OLS) method . After the stepwise backward method we have found a model with 18 terms given in Table \ref{Tab3}. We observe that the obtained model was the same as that found by  \cite{NaesFaergestadCornell}. 

\def\tablename{Table}
\begin{table} [!htp]
\scalefont{0.75}
\begin{center}
\caption{Regression coefficients for the mean model by OLS method} \vspace{0.3cm}
\begin{tabular}{lccccc} \hline
Terms                              &  Estimate         &  Std. Error   &  t value       & p-value          \\ \hline
$x_{1}$                            &   484.624         &   6.363       &  76.161        & 0.0000           \\ 
$x_{2}$                            &   474.875         &  13.369       &  35.521        & 0.0000           \\
$x_{3}$                            &   436.381         &  64.837       &   6.730        & 0.0000           \\
$x_{1}x_{3}$                       &   468.313         & 164.234       &   2.851        & 0.0057           \\ 
$x_{1}x_{2}(x_{1}-x_{2})$          &   375.341         &  94.623       &   3.397        & 0.0002           \\
$x_{1}x_{3}(x_{1}-x_{3})$          &  -403.031         & 199.679       &  -2.018        & 0.0473           \\
$x_{1}z_{1}$                       &    16.768         &   5.452       &   3.076        & 0.0029           \\    
$x_{3}z_{1}$                       &    51.876         &   8.406       &   6.171        & 0.0000           \\ 
$x_{1}x_{2}(x_{1}-x_{2})z_{1}$     &  -144.553         &  60.706       &  -2.381        & 0.0199           \\ 
$x_{1}z_{2}$                       &    54.933         &   6.703       &   8.195        & 0.0000           \\   
$x_{2}z_{2}$                       &    42.504         &   8.470       &   5.018        & 0.0000           \\   
$x_{1}x_{3}z_{2}$                  &   188.762         &  25.167       &   7.500        & 0.0000           \\ 
$x_{1}x_{3}(x_{1}-x_{3})z_{2}$     &  -202.822         &  61.681       &  -3.288        & 0.0016           \\ 
$x_{2}z^{2}_{1}$                   &   -52.644         &  14.972       &  -3.516        & 0.0008           \\ 
$x_{3}z^{2}_{1}$                   &   164.077         &  79.249       &   2.070        & 0.0420           \\ 
$x_{1}x_{3}z^{2}_{1}$              &  -600.046         & 199.173       &  -3.013        & 0.0036           \\ 
$x_{1}x_{2}(x_{1}-x_{2})z^{2}_{1}$ &  -440.721         & 109.730       &  -4.016        & 0.0001           \\ 
$x_{1}x_{3}(x_{1}-x_{3})z^{2}_{1}$ &   525.480         & 244.486       &   2.149        & 0.0349           \\ \hline                     
\end{tabular}
\label{Tab3}
\end{center}
\end{table}

In our analysis, in order to make inference for both models of mean and dispersion, following the algorithm shown in Appendix B, we create a package implemented in the R system for statistical computing \cite{RTeam}. However, other software could be used to obtain these models. 

For comparison between the competitors joint models, based on the ideas of \cite{DarongZhongzhan}, we consider a kind of quasi Akaike's Information Criterion given by $AICq=-2Q_{A}^{+} +2(p+q)$, where $p$ and $q$ are the number of parameters in the models of the mean and dispersion, respectively. The best joint model is one that has the lowest value of $AICq$. A global measure of variation explained by the fitted joint model can be obtained by calculating the pseudo $R^2$ ($R_p^2$), defined as the square of the sample correlation coefficient between $\eta$ and $\varphi(y)$. We can observe that $0 \leq R_p^2 \leq 1$ and the closer to 1 it is, the greater the correlation between the linear predictor and the transformed response observed.

For the starting joint model ($JM_0$), we consider as a plausible alternative to use the same linear predictor for both the mean model ($M_0$) and dispersion  model ($D_0$), that is $\eta_{_{M_0}}=\xi_{_{D_0}}$, whose terms are given in Table \ref{Tab3}. However, other alternatives could be considered for the choice of the linear predictors for the mean and dispersion models, see comments in \cite{LeeNelderI}. The procedure for selecting the best model for the mean and dispersion came from an adaptation of the stepwise backward method, where the models for the mean and dispersion were changed alternately. The Table \ref{Tab4} shows the values of $AICq$ and $R_{p}^{2}$ for some joint models considered. In Table \ref{Tab4}, the linear predictor for the mean model $M_j$ is the linear predictor for the mean model $M_i$ removed the terms in braces, that is, $\eta_{_{M_j}}=\eta_{_{M_i}} -\{terms\}$. The linear predictor for the dispersion model is obtained analogously. We note that when terms are removed from models of the mean and dispersion the values of $AICq$ and $R_{p}^{2}$ are get worse. The best joint model considered is $JM_2$ with $\eta_{_{M_2}}=\eta_{_{M_0}}$, $\xi_{_{D_2}}=\xi_{_{D_0}}-\{x_1z_1, \; x_2z_{1}^{2}\}$,  $AICq=809.8640$ and $R_{p}^{2}=0.9199$.

In order to verify that the model $JM_2$ exhibits non-constant variance, we employ the studentized Breusch and Pagan test \cite{BreuschPagan} which is available in R in the lmtest package \cite{ZeileisHothorn}. The value provided by the test is 26.1733 with p-value = 0.03624, indicating, in fact, that the variance is not constant.

\def\tablename{Table}
\begin{table} [htp!] 
\scalefont{0.65}
\begin{center}
\caption{Values of $AICq$ and $R_{p}^{2}$ for comparison of joint models} \vspace{0.2cm}
\begin{tabular}{llcc} \hline 
Joint model             & Linear predictor$^{\dag}$                                  &  $AICq$                    &  $R_{p}^{2}$                   \\ \hline
\multirow{2}{*}{$JM_0$} & $\eta_{_{M_0}}$ {\tiny (terms given in Table \ref{Tab3})}  & \multirow{2}{*} {813.8589} & \multirow{2}{*} {0.9199}       \\
                        & $\xi_{_{D_0}}$  {\tiny (terms given in Table \ref{Tab3})}  &                            &                                \\ \hline
\multirow{2}{*}{$JM_1$} & $\eta_{_{M_1}}=\eta_{_{M_0}}$                              & \multirow{2}{*} {826.3424} & \multirow{2}{*} {0.8995}       \\
                        & $\xi_{_{D_1}}=\xi_{_{D_0}} -\{x_1z_1\} $                   &                            &                                \\ \hline
\multirow{2}{*}{$JM_2$} & $\eta_{_{M_2}}=\eta_{_{M_0}}$                              & \multirow{2}{*} {809.8640} & \multirow{2}{*} {0.9199}       \\
                        & $\xi_{_{D_2}}=\xi_{_{D_0}}-\{x_1z_1, \; x_2z_{1}^{2}\}$    &                            &                                \\ \hline
\multirow{2}{*}{$JM_3$} & $\eta_{_{M_3}}=\eta_{_{M_1}} -\{x_1x_2(x_1-x_2)z_1\}$      & \multirow{2}{*} {819.3661} & \multirow{2}{*} {0.8995}       \\
                        & $\xi_{_{D_3}}=\xi_{_{D_0}}-\{x_3z_1, \; x_1x_2(x_1-x_2)z_1, \; x_1x_3z_2, \; x_1x_3(x_1-x_3)z_2\}$ &        &            \\ \hline
\multicolumn{4}{l}{$^{\dag}$\tiny{$\eta_{_{M_j}}=\eta_{_{M_i}} -\{terms\}$ is the linear predictor $\eta_{_{M_i}}$} removed the terms in braces.}  \\
\multicolumn{4}{l}{$^{\dag}$\tiny{$\xi_{_{D_j}}=\xi_{_{D_i}} -\{terms\}$ is the linear predictor $\xi_{_{D_i}}$} removed the terms in braces.}     \\
\end{tabular}
\label{Tab4}
\end{center}
\end{table}


Note that the extended quasi-likelihood is based on a saddlepoint approximation to an exponential family likelihood, i.e., the GLM family, as point out by \cite{LeeNelderPawitan}, p. 81. Thus, we can build an approximate likelihood ratio test. Suppose we are considering two nested (extended quasi likelihood) models EQL1, with $n_1$ parameters, and EQL2, with $n_2$ parameters, such that $EQL1 \subset EQL2$ and therefore $n_2 > n_1$. Let $-2Q_{A1}^{+}$ and $-2Q_{A2}^{+}$ be the adjusted extended quasi likelihood, given in equation (\ref{eq4}), for the two models respectively. As well as $\ln L \approx Q_{A}^{+}$, where $L$ is the likelihood function, the approximate likelihood test is given by $-2(Q_{A1}^{+}-Q_{A2}^{+}) \sim \chi^{2}_{n_2 - n_1}$. In our analysis, we consider the dispersion model as fixed and the test is built only for the terms in the mean model. Thus, we are considering EQL1 as a full model with $p+q$ parameters, that is, $p$ parameters for the mean model and $q$ parameters for the dispersion model. The EQL2 model is the EQL1 model with a term less in the mean model, with $p+q-1$ parameters. Thus, the approximate likelihood test is given by $\lambda = -2(Q_{Ax}^{+}-Q_{A}^{+}) \sim \chi^{2}_{1}$, where $-2Q_{A}^{+}$ is the adjusted extended quasi likelihood for the full model with $p+q$ parameters, $-2Q_{Ax}^{+}$ is the adjusted extended quasi likelihood for the full model removed the term $x$, with $p+q-1$ parameters and ${\cal{X}}^{2}_1$ represents the chi-square distribution with one degree of freedom.    

For the dispersion model we use the analysis of deviance, because in this case we are assuming a true GLM. The analysis is conducted considering the model for mean as fixed, that is, for each iteration of the algorithm, shown in Appendix B, the analysis of deviance is performed after the mean model has been obtained. The statistic $D^d_x - D^d \sim {\cal{X}}^{2}_1$, where $D^d$ is the deviance for a dispersion model with $q$ parameters, $D^d_x$ is the deviance for a dispersion model removed the term $x$ with $q-1$ parameters.  The deviance for the dispersion model is given by $\sum_{i=1}^{n}d_{d_i}$, where $d_{d_i}$ is the deviance component given in Table \ref{Tab1}. 

Note that we could have used the approximate likelihood ratio test to analyse the terms of the dispersion model, but, as we are assuming the model for dispersion as known, we prefer to use the deviation analysis. Also note that the criterion of approximate information could be used in both mean and dispersion analysis, but this was not carried out. Finally, we consider the Wald's method in both mean and dispersion analysis.

The tests for significance of parameters in the models of the mean and dispersion in the joint model $JM_2$ are given in Tables \ref{Tab5} and \ref{Tab6}, respectively.

\def\tablename{Table}
\begin{table} [htp!] 
\scalefont{0.65}
\begin{center}
\caption{Regression coefficients for the mean model in JMMD} \vspace{0.3cm}
\begin{tabular}{lrrrrrrrr} \hline 
\multicolumn{3}{}{ } & \multicolumn{2}{c}{ Wald Method} &  & \multicolumn{3}{c}{EQD Method $^{\dag}$}     \\ \cline{4-5} \cline{7-9}
Terms                              &  Estimate   &  Std. Error   &  t value   & p-value  &  & $-2Q_{Ax}^{+}$  &   Chi-Sq. value  & p-value \\ \hline
$x_{1}$                            &   482.801   &   6.225       &  77.562    & 0.0000   &  & 1023806.141     &  1023064.278 &   0.0000  \\ 
$x_{2}$                            &   470.863   &  11.477       &  41.026    & 0.0000   &  &  217854.278     &   217112.415 &   0.0000  \\
$x_{3}$                            &   437.682   &  21.961       &  19.930    & 0.0000   &  & 1045693.586     &  1044951.723 &   0.0000  \\
$x_{1}x_{3}$                       &   488.284   &  45.242       &  10.793    & 0.0000   &  &  260668.354     &   259926.491 &   0.0000  \\
$x_{1}x_{2}(x_{1}-x_{2})$          &   247.959   &  90.910       &   2.728    & 0.0080   &  &    1706.425     &      964.562 &   0.0000  \\ 
$x_{1}x_{3}(x_{1}-x_{3})$          &  -302.267   &  87.106       &   3.470    & 0.0009   &  &    2820.485     &     2078.622 &   0.0000  \\
$x_{1}z_{1}$                       &    14.276   &   4.965       &   2.875    & 0.0053   &  &     757.165     &       15.302 &   0.0001  \\    
$x_{3}z_{1}$                       &    52.470   &   6.782       &   7.737    & 0.0000   &  &     839.941     &       98.078 &   0.0000  \\ 
$x_{1}x_{2}(x_{1}-x_{2})z_{1}$     &  -138.624   &  47.359       &   2.927    & 0.0046   &  &     761.398     &       19.535 &   0.0000  \\ 
$x_{1}z_{2}$                       &    57.738   &   5.930       &   9.736    & 0.0000   &  &   14109.226     &    13367.363 &   0.0000  \\   
$x_{2}z_{2}$                       &    52.242   &   4.648       &  11.240    & 0.0000   &  &    3290.771     &     2548.908 &   0.0000  \\   
$x_{1}x_{3}z_{2}$                  &   154.184   &  12.388       &  12.447    & 0.0000   &  &   24544.717     &    23802.854 &   0.0000  \\ 
$x_{1}x_{3}(x_{1}-x_{3})z_{2}$     &  -281.902   &  18.707       &  15.069    & 0.0000   &  &    2396.114     &     1654.251 &   0.0000  \\ 
$x_{2}z^{2}_{1}$                   &   -42.406   &  13.145       &   3.226    & 0.0019   &  &    833.364      &       91.501 &   0.0000  \\ 
$x_{3}z^{2}_{1}$                   &   143.488   &  51.188       &   2.803    & 0.0065   &  &   1456.674      &      714.811 &   0.0000  \\ 
$x_{1}x_{3}z^{2}_{1}$              &  -565.182   & 130.336       &   4.336    & 0.0000   &  &   1750.571      &     1008.708 &   0.0000  \\ 
$x_{1}x_{2}(x_{1}-x_{2})z^{2}_{1}$ &  -330.179   & 102.329       &   3.227    & 0.0019   &  &    847.981      &      106.118 &   0.0000  \\ 
$x_{1}x_{3}(x_{1}-x_{3})z^{2}_{1}$ &   362.238   & 168.999       &   2.321    & 0.0231   &  &    814.162      &       72.299 &   0.0000  \\ \hline 
\multicolumn{9}{l}{$^{\dag}$ \scriptsize{$-2Q^{+}_{Ax}$ is the value of the EQD for a joint model with fixed dispersion model and with the mean model}}\\
\multicolumn{9}{l}{\scriptsize{removed the term $x$, with $n_{p}-1=17$ parameters. $-2Q^{+}_{A}=741.863$ with $n_{p}=18$ parameters.}}
             
\end{tabular}
\label{Tab5}
\end{center}
\end{table}

\def\tablename{Table}
\begin{table} [!htp]
\scalefont{0.65}
\begin{center}
\caption{Regression coefficients for the dispersion model in JMMD} \vspace{0.3cm}
\begin{tabular}{lrrrrrrrr} \hline
\multicolumn{3}{}{ } & \multicolumn{2}{c}{ Wald Method} &  & \multicolumn{3}{c}{Deviance Method $^{\ddag}$}     \\ \cline{4-5} \cline{7-9}
Terms                              &  Estimate   &  Std. Error   &  t value   & p-value  & &  $D^{d}_{x}$  & Chi-Sq. value & p-value \\ \hline
$x_{1}$                            &     6.028   &   0.498       &   12.096   & 0.0000   & &    12047.685  &    11892.898  & 0.0000  \\ 
$x_{2}$                            &     5.221   &   0.708       &    7.372   & 0.0000   & &     1248.478  &     1093.691  & 0.0000  \\
$x_{3}$                            &    18.488   &   5.799       &    3.188   & 0.0021   & &     126.55e5  &     126.55e5  & 0.0000  \\
$x_{1}x_{3}$                       &   -47.758   &  15.743       &   -3.034   & 0.0033   & &      856.731  &      701.944  & 0.0000  \\
$x_{1}x_{2}(x_{1}-x_{2})$          &    25.977   &   7.414       &    3.504   & 0.0008   & &      324.438  &      169.651  & 0.0000  \\ 
$x_{1}x_{3}(x_{1}-x_{3})$          &    52.270   &  18.295       &    2.857   & 0.0056   & &     2153.342  &     1998.555  & 0.0000  \\
$x_{3}z_{1}$                       &    -0.290   &   0.567       &   -0.511   & 0.6111   & &      155.407  &        0.620  & 0.4310  \\
$x_{1}x_{2}(x_{1}-x_{2})z_{1}$     &    -5.249   &   4.775       &   -1.099   & 0.2752   & &      159.466  &        4.679  & 0.0305  \\ 
$x_{1}z_{2}$                       &    -0.456   &   0.537       &   -0.849   & 0.3986   & &      160.094  &        5.307  & 0.0212  \\   
$x_{2}z_{2}$                       &     1.846   &   0.706       &    2.615   & 0.0108   & &      185.159  &       30.372  & 0.0000  \\   
$x_{1}x_{3}z_{2}$                  &    -1.592   &   2.078       &   -0.766   & 0.4461   & &      157.511  &        2.724  & 0.0989  \\ 
$x_{1}x_{3}(x_{1}-x_{3})z_{2}$     &    -6.438   &   5.222       &   -1.233   & 0.2216   & &      158.718  &        3.931  & 0.0474  \\ 
$x_{3}z^{2}_{1}$                   &   -15.316   &   6.456       &   -2.372   & 0.0203   & &      541.654  &      386.866  & 0.0000  \\ 
$x_{1}x_{3}z^{2}_{1}$              &    56.251   &  17.325       &    3.247   & 0.0018   & &     182.66e5  &     182.66e5  & 0.0000  \\ 
$x_{1}x_{2}(x_{1}-x_{2})z^{2}_{1}$ &   -32.718   &   8.537       &   -3.833   & 0.0003   & &      435.597  &      280.810  & 0.0000  \\ 
$x_{1}x_{3}(x_{1}-x_{3})z^{2}_{1}$ &   -51.608   &  20.485       &   -2.519   & 0.0139   & &     1396.891  &     1242.104  & 0.0000  \\ 
\hline 
\multicolumn{9}{l}{$^{\ddag}$ \scriptsize{$D^{d}_{x}$ is the value of the deviance for the dispersion model removed the term $x$, with $n_{q}-1=17$ parameters}}.\\
\multicolumn{9}{l}{\scriptsize{$D^{d}_{A}=154.787$, with $n_{q}=18$ parameters. In both $D^{d}$ and $D^{d}_{x}$ the mean model is the same.}}                     
\end{tabular}
\label{Tab6}
\end{center}
\end{table}

\begin{figure}[!htb]
\centering
\includegraphics[scale=0.75]{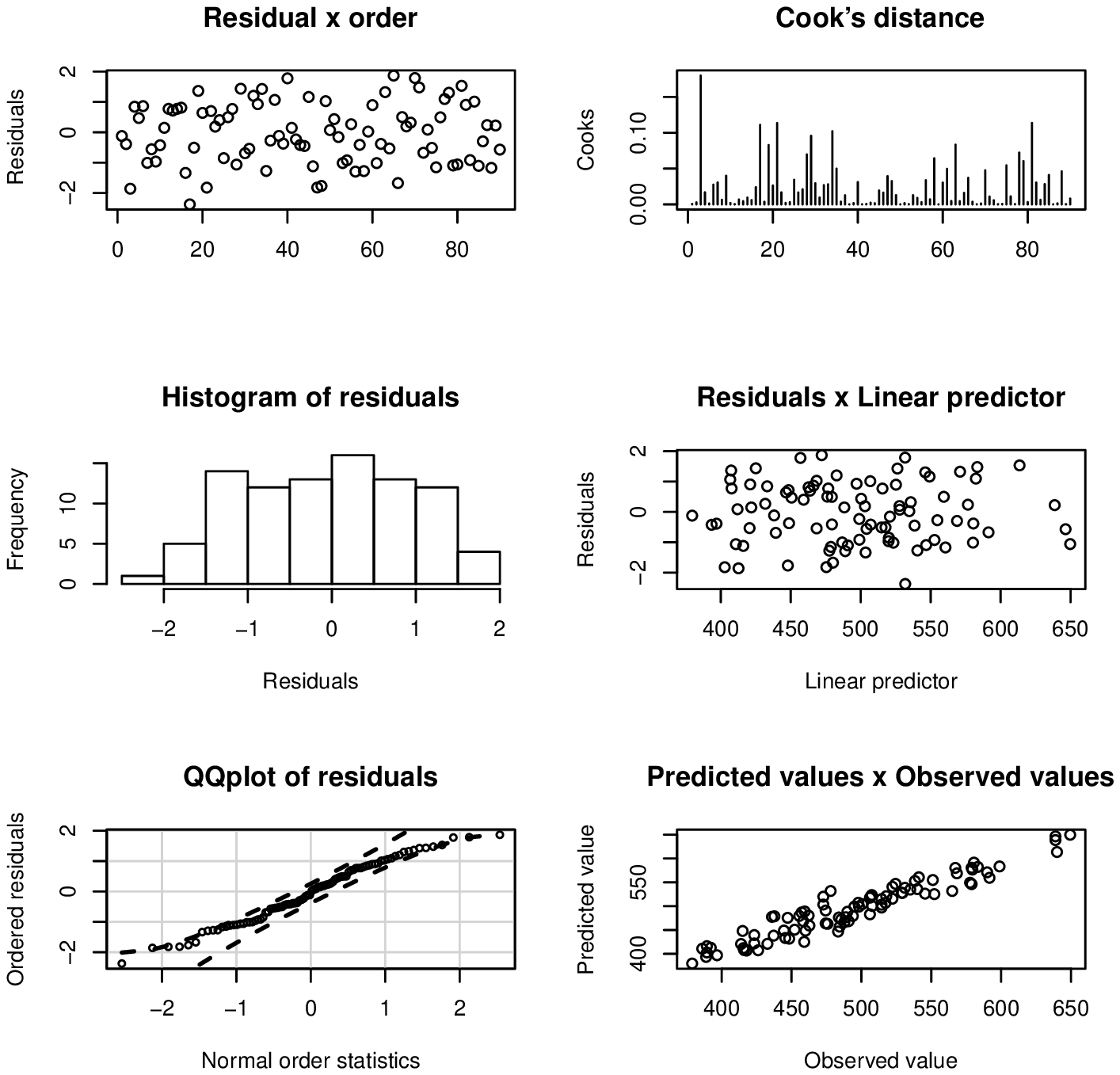}
\caption{ \scriptsize Six diagnostic plots for data of bread mixture - mean model. The upper left panel plots standardized deviance residuals against the order of observations, the upper right panel plots the Cook's distances versus the order of observations, the middle left panel displays the diagonal elements of the matrix hat against the predicted values, the middle right panel plots the standardized deviance residuals versus the linear predictor, the lower left panel displays the half-normal plot of absolute standardized deviance residuals with a simulated envelope and the lower right panel presents the predicted values versus observed values.}
\label{fig2}
\end{figure}

\begin{figure}[!htb]
\centering
\includegraphics[scale=0.75]{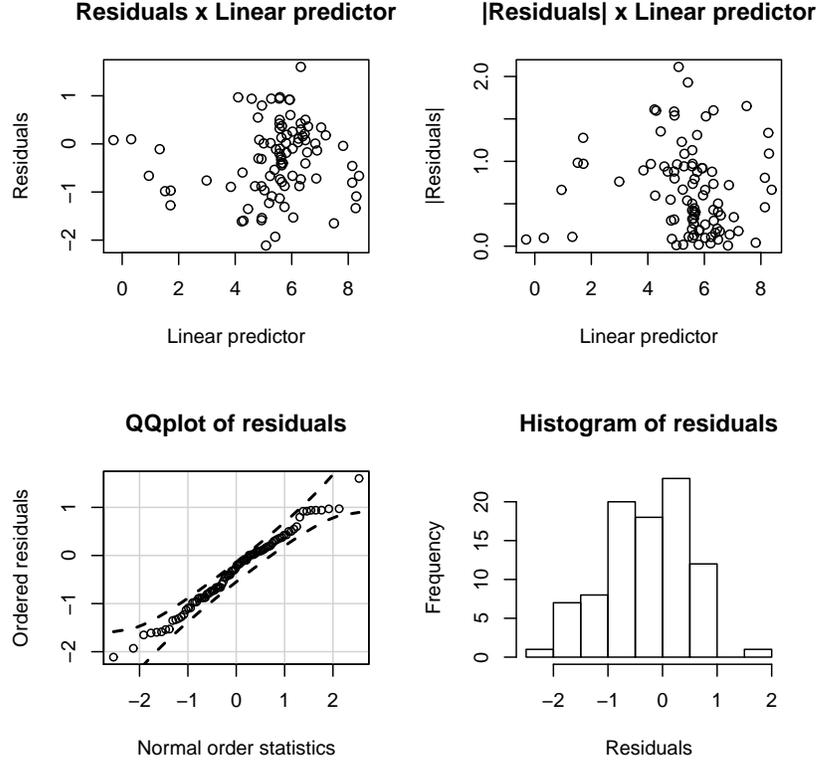}
\caption{\scriptsize Four diagnostic plots for data of bread mixture - dispersion model. The upper left panel plots standardized deviance residuals against the linear predictor, the upper right panel plots the absolute standardized deviance residuals versus the linear predictor, the lower left panel displays the normal plot of residuals with a simulated envelope and the lower right panel presents the histogram of standardized deviance residuals.}
\label{fig3}
\end{figure}

As pointed out by \cite{LeeNelderPawitan}, for the residual analysis in the mean model we use the standardized deviance residual, given by $r^m_i=sign(y_i - \hat{\mu}_i)\sqrt{\frac{d^{*}_{i}}{\phi_i}}$ and  $r^d_i=sign(d_i - \hat{\phi}_i)\sqrt{\frac{d^{*}_{d_i}}{\hat{\phi}}}$, for the residual analysis in the dispersion model, with $\hat{\phi}=\frac{\sum_{i=1}^{n}d_{d_i}}{n-q}$ and $d^{*}_{d_i}=\frac{d_{d_i}}{1-h_{d_i}}$, where $h_{d_i}$ is the $i$th element of the diagonal of $\mat{V}^{\frac{1}{2}}\mat{U}(\mat{U}^t\mat{V}\mat{U})^{-1}\mat{U}\mat{V}^{\frac{1}{2}}$ (see Appendix A). The goodness of fit of the mean and dispersion models is assessed using different types of diagnostic displays shown in the Figures \ref{fig2} and \ref{fig3} respectively.

The final model for mean is given by

{\small
\begin{eqnarray}  \label{eq9} 
E(Y|Z_{1},Z_{2}) & = & 482.80x_{1} + 470.86x_{2} + 437.68x_{3} + 488.28x_{1}x_{3} + 247.96x_{1}x_{2}(x_1-x_2) - \nonumber \\ 
                 &   & 302.27x_{1}x_{3}(x_1-x_3)+ [14.28x_1 + 52.47x_3 - 138.62x_1x_2(x_1-x_2)]z_1 + \nonumber \\
                 &   & [57.74x_1 + 52.24x_2 + 154.18x_1x_3 -281.90x_1x_3(x_1 - x_3)]z_2 + \nonumber \\
                 &   & [-42.41x_2 + 143.49x_3 - 565.18x_1x_3 -330.18x_1x_2(x_1 - x_2) + \nonumber \\
                 &   & 392.24x_1x_3(x_1 - x_3)]z_{1}^{2} 
\end{eqnarray} 
}
and the model for dispersion is the same model for the variance, i.e., $Var(Y|Z_1,Z_2)$ $=\phi$, because $V(\mu)=1$, and it is given by

{\small
\begin{eqnarray} \label{eq10}
Var(Y|Z_{1},Z_{2}) & = & exp \{6.03x_{1} + 5.22x_{2} + 18.49x_{3} -47.76x_{1}x_{3} + \nonumber \\ 
                   &   & 25.98x_1x_2(x_1 - x_2) + 52.27x_1x_3(x_1 - x_3) -5.25x_1x_2(x_1 - \nonumber \\
                   &   & x_2)z_1 + [-0.46x_1 + 1.84x_2]z_2 + [-15.32x_3 + 56.25x_1x_3 -  \nonumber \\
                   &   & 32.72x_1x_2(x_1 -x_2) -51.61 x_1x_3(x_1 - x_3)]z_{1}^{2}\}
\end{eqnarray}
}

To obtain the expressions for $E(Y)$ and $Var(Y)$, we supposed that the coded noise variables $Z_1\sim N(\mu_1,\sigma^{2}_{1})$, $Z_2\sim N(\mu_2,\sigma^{2}_{2})$ and are independent random variables. Now, knowing that $E(Y)=E(E(Y|Z_{1},Z_{2}))$ we have that 
{\small
\begin{eqnarray} \label{eq11}
E(Y)& = &482.80x_1 + 470.86x_2 + 437.68x_3 + 488.28x_1x_3 + 247.96x_{1}x_{2}(x_1-x_2) - \nonumber \\
    &   &302.27x_{1}x_{3}(x_1-x_3)+ [14.28x_1 + 52.47x_3 - 138.62x_1x_2(x_1-x_2)]\mu_1 +   \nonumber \\ 
    &   &[57.74x_1 + 52.24x_2 + 154.18x_1x_3 -281.90x_1x_3(x_1 - x_3)]\mu_2 + [-42.41x_2 +   \nonumber \\
    &   &143.49x_3 - 565.18x_1x_3 -330.18x_1x_2(x_1 - x_2) + 392.24x_1x_3(x_1 - x_3)] (\mu_1^{2} + \nonumber \\
    &   &\sigma_1^{2}),
\end{eqnarray}
}
\noindent where $E(Z_1^{2})=\sigma_1^{2} + \mu_1^{2}$.  Using the fact that $Var(Y)=E(Var(Y|Z_{1},Z_{2}))+Var(E(Y|Z_{1},$ $Z_{2}))$ and that if $Z \sim N(\mu,\sigma^{2})$ then $E(e^{\alpha_1Z})=e^{\alpha_1\mu +(\alpha_1^{2}\sigma^{2})/2}$, $E(e^{\alpha_1Z + \alpha_2Z^{2}}) =\left(1/\sqrt{k_2}\right)$ $e^{\frac{1}{2\sigma^{2}}\left(\frac{k_1^{2}}{k_2} - \mu^{2}\right)}$, with $k_2 > 0$, and $Var(\alpha_1Z + \alpha_2Z^{2})$ $=\sigma^{2}(\alpha_1^{2} + 2\alpha_2^{2}(2\mu^{2} + \sigma^{2}) + 4\alpha_1\alpha_2\mu)$, where $k_1=\alpha_1\sigma^{2} + \mu$, $k_2=1-2\sigma^{2}\alpha_2$, and $\alpha_1$, $\alpha_2$ are constant, we have that

\begin{eqnarray} \label{eq12}
Var(Y)& = & E(e^{m_1}e^{m_2Z_1}e^{m_3Z_2}e^{m_4Z_{1}^{2}}) + Var(m_5Z_1+ m_6Z_2 + m_7Z_1^{2}) = \nonumber \\
      &   & e^{m_1}e^{m_3\mu_2 + \frac{m_3^{2}\sigma_2^{2}}{2}}\frac{1}{\sqrt{1-2\sigma_1^{2}m_4}}e^{\frac{1}{2\sigma_{1}^{2}}\left(\frac{(m_2\sigma_1^{2} + \mu_1)^{2}} {1-2\sigma_1^{2}m_4} - \mu_1^{2}\right)} + m_6^{2}\sigma_2^{2} + \nonumber \\
    &   & \sigma_1^{2}(m_5^{2} + 2m_7^{2}(2\mu_1^{2} + \sigma_1^{2}) + 4m_5m_7\mu_1),
\end{eqnarray}

\noindent where $m_1=6.03x_{1} + 5.22x_{2} + 18.49x_{3} -47.76x_{1}x_{3} + 25.98x_1x_2(x_1 - x_2) + 52.27x_1x_3(x_1 - x_3)$, $m_2=-5.25x_1x_2(x_1 - x_2)$, $m_3=-0.46x_1 + 1.84x_2$, $m_4=-15.32x_3 + 56.25x_1x_3 -32.72x_1x_2(x_1 -x_2) -51.61 x_1x_3(x_1 - x_3)$, $m_5=52.47x_3$, $m_6=57.74x_1 + 52.24x_2 + 154.18x_1x_3 -281.90x_1x_3(x_1 - x_3)$, $m_7=-42.41x_2 + 143.49x_3 - 565.18x_1x_3 -330.18x_1x_2$ $(x_1 - x_2) +  392.24x_1x_3 (x_1 - x_3)$. 

\subsection{Modeling of mean and variance via delta method}

The use of the delta method to obtain models for the mean and variance in mixture experiments with process variables was introduced by \cite{Goldfarb}. Goldfarb et. al  \cite{Goldfarb} consider the model for the response variable $Y$ given by $Y=\eta(\mat{x},\mat{w},\mat{z})+ \varepsilon$, where $\mat{x}$ is the vector of the mixture components, $\mat{w}$ is the vector of the controllable process variables, $\mat{z}$ is the vector of the noise variables, $\varepsilon$ is the random error with $\varepsilon \sim N(0,\sigma^2)$ and $\eta(\mat{x},\mat{w},\mat{z})$ is the linear predictor, containing quadratic or special cubic mixture terms, interactions between the mixture components and the controllable process variables, interactions between mixture components and the noise variables, and interactions among all three types of variables. Considering the noise variables $\mat{Z}^t=(Z_1,\ldots,Z_b)$, they assume that $E(\mat{Z})=\mat{0}$ and $Var(\mat{Z})=Diag(\sigma^{2}_{1}, \ldots,\sigma^{2}_{b})$ is a $b \times b$ diagonal matrix with the variances of the noise variables on the diagonal. Goldfarb et al. \cite{Goldfarb} obtain $E(Y)$ and $Var(Y)$ via delta method considering a first-order Taylor series approximation of the model around the mean of $\mat{Z}$, taken as zero.

\subsubsection{Model building strategy} \vspace{0.3cm}
\label{mbs-delta-method}

As proposed in Section \ref{delta-method}, the delta method will be applied considering the second-order expansion of Taylor series around the mean of noise variables. For we apply the delta method we consider the final model obtained by \cite{NaesFaergestadCornell} whose terms are shown in Table \ref{Tab3}. Thus, the delta method is applied to the estimated model, given by $\hat{y}=\eta(\mat{x},\mat{z})=c_{1} + c_{2}z_{1} + c_{3}z_{2} + c_{4}z_{1}^{2}$, where $c_{1}=484.62x_1 + 474.88x_2 + 436.38x_3 + 468.31x_1x_3 + 375,34x_1x_2(x_1-x_2) - 403.03x_1x_3(x_1-x_3)$, $c_{2}=16.77x_{1} + 51.88x_3 - 144.55x_{1}x_{2}(x_{1}-x_{2})$, $c_{3}=54.93x_{1} + 42.50x_{2} + 188.76x_{1}x_{3} - 202.82x_{1}x_{3}(x_{1}-x_{3})$ and $c_{4}= -52.64x_{2} + 164.08x_{3} - 600.05x_{1}x_{3} - 440.72x_{1}x_{2}(x_{1} - x_{2}) + 525.48x_{1}x_{3}(x_{1}-x_{3})$ represent the constant terms in relation to $\mat{z}$. 

The response model assumed by \cite{NaesFaergestadCornell} is  $y=\eta(\mat{x},\mat{z})+ \varepsilon$ with $\varepsilon \sim N(0,\sigma^2)$, where $\mat{z}^t=(z_{1},z_{2})$. Thus, considering $E(\mat{Z})=\mat{\mu}^{t}=(\mu_{1},\mu_{2})$ and $Var(\mat{Z})=Diag(\sigma^{2}_{1},\sigma^{2}_{2})$, from Section \ref{delta-method} we have that $\eta'\left(\mat{x},\mat{\mu}\right)=\left.\displaystyle\frac{\partial}{\partial \mat{z}} \eta\left(\mat{x},\mat{z}\right)\right|_{\mat{z}=\mat{\mu}}=\left(
\begin{array}{c}
c_2 + 2c_{4}\mu_{z_1} \\
c_{3}                 \\
\end{array}
\right)$ and $\mat{\cal{H}}=\left.\displaystyle\frac{\partial^{2}}{\partial \mat{z}\partial \mat{z}^t} \eta \left(\mat{x},\mat{z}\right)\right|_{\mat{z}=\mat{\mu}}=$ $\left(
\begin{array}{cc}
2c_4 & 0 \\
0    & 0 \\
\end{array}
\right)$.

Now, from the equation (\ref{eq5}) we obtain $y=\eta(\mat{x},\mat{z})+ \varepsilon = c_{1}+c_{2}\mu_{1}+c_{3}\mu_{2}+c_{4}\mu^{2}_{1}+(z_{1}-\mu_{1})(c_{2}+2c_{4}\mu_{1})+c_{3}(z_{2}-\mu_{2})+2c_{4}(z_{1}-\mu_{1})^2 + \cal{R} + \varepsilon$.

Thereafter, using the fact that if $Z\sim N(\mu,\sigma^2)$ then $E(Z^3)=\mu^{3}+3\sigma^{2}\mu$, $Var(Z^2)=4\sigma^{2}\mu^{2}+2\sigma^{4}$ and $Var(aZ+bZ^2)=a^2Var(Z)+b^2Var(Z^2)-2ab[E(Z^3)-E(Z)E(Z^2)]$, the models for mean and variance can be obtained from equations (\ref{eq6}) and (\ref{eq7}), respectively, and are given by 

\begin{equation} \label{eq13}
E(Y)=c_{1} + c_{2}\mu_{1} +c_{3}\mu_{2} + c_{4}\mu^{2}_{1} + 2c_{4}\sigma^{2}_{1} 
\end{equation}

and 

\begin{equation} \label{eq14}
Var(Y)=\left(c_{2}^{2} + 4c_{2}c_{4}\mu_{1} +4c_{4}^{2}\mu_{1}^{2} \right)\sigma_{1}^{2} + 8c_{4}^{2}\sigma^{4}_{1} + c_{3}^{2}\sigma_{2}^{2} + \sigma^{2}. 
\end{equation}

\subsection{Optimization process} \vspace{0.3cm}
\label{Optimization_process}

Following the Taguchi's idea for the quality improvement, see \cite{Taguchi}, after we found the equations for $E(Y)$ and  $Var(Y)$,  we have to solve the following minimization problem

\begin{eqnarray} \label{eq15} 
   Min \; Var(Y) &  \nonumber \\
   Subject \; to & \left\{ \begin{array}{l}               
                           E(Y)=530.0  \\
                           x_{1}+x_{2}+x_{3}=1.0 \\
                           0.25 \le x_{1} \le 1.0 \\
                           0.0 \le x_{2} \le 0.75 \\
                           0.0 \le x_{3} \le 0.75  
                           \end{array} \right.                             
 \end{eqnarray}

\noindent where $E(Y)$ and $Var(Y)$ are functions of $x_{1}$, $x_{2}$, $x_{3}$, $\mu_1$, $\mu_2$, $\sigma_{1}^{2}$ and $\sigma_{2}^{2}$. Fort the JMMD the expressions for $E(Y)$ and $Var(Y)$ are given by the equations (\ref{eq11}) and (\ref{eq12}), respectively, and we still must assume an additional constraint, i.e., $1-2\sigma_1^{2}m_4 > 0$. For the delta method the expressions for $E(Y)$ and $Var(Y)$ are given in equations (\ref{eq13}) and (\ref{eq14}), respectively, and we use $\hat{\sigma}^{2}=\frac{D}{n-p}=58.36$ as an estimative for $\sigma^{2}$ in equation (\ref{eq14}), where $D=4201,615$ is the deviance for the model shown in Table \ref{Tab3}, with $n=90$ observations and $p=18$ parameters.

We solve the optimization problem, considering various scenarios involving the values of mean and variance for the random variables mixing time and proofing time. The Table \ref{Tab7} shows, for the JMMD and the delta method, the optimum combination for the mixture and its variance estimated for each scenario involving the noise variables.

\def\tablename{Table}
\begin{table} [htp!] 
\scalefont{0.70}
\begin{center}
\caption{Optimal values of mixture for various scenarios involving the noise variables, where the models for mean and variance were obtained via JMMD and delta method.} \vspace{0.3cm}
\begin{tabular}{ccccccc} \hline 
\multicolumn{2}{}{ } & \multicolumn{2}{c} {JMMD}   &  &\multicolumn{2}{c} {Delta Method}  \\ \cline{3-4}  \cline{6-7}
Mix. Time $^{\dag}$        &  Proof. Time $^{\ddag}$    &    Optimum         &  Variance   &    & Optimum          & Variance    \\ 
($\mu_{m},\sigma_{m}^{2}$) & ($\mu_{p},\sigma_{p}^{2}$) &    ($x_1,x_2,x_3$) &  estimated  &    & ($x_1,x_2,x_3$)  & estimated   \\ \hline
$(10.0,6.25)$  & $(47.50,9.766)$    &  (0.303, 0.483, 0.214)&  160.916 &   &(0.250, 0.063, 0.687)& 691.387  \\
$(12.5,6.25)$  & $(44.375,9.766)$   &  (0.310, 0.534, 0.156)&  98.893  &   & (0.250, 0.066, 0.684)& 548.857 \\ 
$(15.0,6.25)$  & $(41.25,9.766)$    &  (0.306, 0.568, 0.126)&   66.912 &   &(0.250, 0.037, 0.713)& 473.602  \\ 
$(15.0,25.0)$  & $(47.500,39.063)$  &  (0.300, 0.560, 0.140)&  272.526 &   &(0.250, 0.168, 0.582)& 1569.286 \\
$(15.0,56.25)$ & $(47.50,87.891)$   &  (0.250, 0.541, 0.209)& 1061.528 &   &(0.250, 0.005, 0.745)& 7406.725 \\
$(20.0,6.25)$  & $(53.75,9.766)$    &  (0.298, 0.598, 0.104)&  215.019 &   &(0.250, 0.423, 0.327)& 217.547  \\ 
$(20.0,25.0)$  & $(53.75,39.0625)$  &  (0.286, 0.599, 0.115)&  432.377 &   &(0.250, 0.393, 0.357)& 794.810  \\
$(20.0,56.25)$ & $(53.75,87.891)$   &  (0.439, 0.498, 0.063)& 1497.716 &   &(0.250, 0.326, 0.424)& 2421.067 \\ \hline
\multicolumn{7}{l}{$^{\dag}$\tiny{Mixing time is a normal random variable with mean $\mu_m$ and variance $\sigma_{m}^{2}$.}}      \\ 
\multicolumn{7}{l}{$^{\ddag}$\tiny{Proofing time is a normal random variable with mean \; $\mu_p$ \; and variance  $\sigma_{p}^{2}$. }} \\          
\end{tabular}
\label{Tab7}
\end{center}
\end{table}

From the scenarios considered in Table \ref{Tab7}, we can observe that, for the delta method, the proportion for $x_1$ is not affected by changes in the parameters of the noise variables and that, for all scenarios considered, the variance estimated using the delta method is greater than the variance estimated by the JMMD. Note also that, given the distributions for the random variables mixing time and proofing time, the probability distributions for the variables $Z_1$ and $Z_2$ are obtained by $Z_1 \sim N(\mu_1, \sigma_1^2)$, $Z_2 \sim N(\mu_2,\sigma_2^2)$, where $\mu_1=(\mu_m -15)/10$, $\sigma_1^2=\sigma_m^2/100$, $\mu_2=(\mu_p -47.5)/12.5$ and $\sigma_2^2=\sigma_p^2/156,25$.

\section{Discussion}
\label{finalconsiderations}

In this paper, we have applied the delta method and the joint modeling of mean and dispersion in a mixture problem with noise variables, with the goal of finding an optimal combination of the mixture ingredients, in order to make the mean of the response variable robust to the noise conditions.

In bread-making example, that had as objective to investigate and to value the final quality of flour composed by different mixtures of wheat flour for production of bread, the optimal combination for the mixture of flour should be obtained to be robust to the mixing and proofing times of the dough, considered as noise variables. The results shown in Section \ref{bread-making-problem} give a comprehensive treatment for the problem in each of the approaches considered. 
 
In our analysis we have considered the noise variables as random variables with Gaussian distribution and we considered various scenarios involving the noise variables, shown in Table \ref{Tab7}. We have obtained optimal combinations of mixture that were robust to the noise conditions, i.e., for the mixtures obtained is expected that the bread produced will have a mean volume of 530 ml, independent of the noise situations for which the mixture was exposed. From the scenarios considered in Table \ref{Tab7}, we can also observe that the variance estimated using the delta method is greater than the variance estimated by the JMMD. However, we can not compare the two approaches considered because there is no method for such a comparison. 

In the optimization process, shown in Subsection \ref{Optimization_process}, the variables $Z_1$ and $Z_2$ were considered noise variables, however, in some situations, some process variables can be considered controllable variables, this does not alter the procedures shown in Subsections \ref{Model_building_strategy} and \ref{mbs-delta-method}, nevertheless for the optimization process given in Subsection \ref{Optimization_process}, such variables are not considered random variables and the optimization process can be conducted for fixed values of these variables.

For the bread-making example, we have considered the Gaussian distribution for the noise variables, however other distributions could be considered, for example, the distributions Gamma or lognormal, but the procedure for finding $E(Y)$ and $Var(Y)$ would be harder. We also have considered independence between noise variables, but for situations where this assumption can not be considered, the complexity of the optimization process increases. Dependence on noise variables using the delta method with first order approximation is considered by \cite{Goldfarb} and for the JMMD still needs further study.

The mean model obtained by \cite{NaesFaergestadCornell} was used as base to apply the delta method. The delta method was applied considering the second-order expansion of Taylor series around the mean of noise variables, however, we could also have used a less accurate first-order approximation. It is worth mentioning that in the case of the delta method, there is only an approximation to the variance. That is, there is no statistical model associated with the variance, just as there is in the case of JMMD. 

\begin{figure}[!htb]
\centering
\includegraphics[scale=0.45]{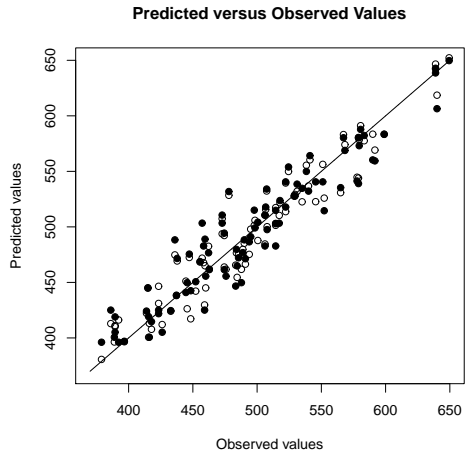}
\caption{\scriptsize Comparison between the two models considered for the mean of response. Filled balls represent the values predicted by our model, given in equation (\ref{eq9}), using the JMMD. Empty balls represent the values predicted by the model obtained by \cite{NaesFaergestadCornell}, whose terms in the linear predictor are shown in Table \ref{Tab3}.}
\label{fig4}
\end{figure}

One can not make formal comparisons between our analysis, using the JMMD, and that proposed by \cite{NaesFaergestadCornell} because they are different approaches and there is no a measure that allows comparisons, however we can use the values of $R_p^2$ and the graphs of the observed values versus the predicted values by the models considered as a criterion for comparison. For the model proposed by \cite{NaesFaergestadCornell}, whose terms in the linear predictor are shown in Table \ref{Tab3}, $R_p^2=0.9189$ and for our model, given in equation (\ref{eq9}), $R_p^2=0.9199$. The Figure \ref{fig4} shows the graphs of the observed values versus the predicted values by both models considered. It should also be mentioned that as much using the delta method as JMMD, we get a model for the variance and it is possible to construct an optimization problem which allows to obtain optimum values  for the ingredients of the mixture making the mean response robust to noise factors. 

It is important to emphasize that when we use JMMD other distributions, that not only the Gaussian distribution, could be considered for the mean model, for instance, distributions for counts or proportions. However, for example, in case of a model for the mean is of Poisson type, with $V(\mu)=\mu$, the complexity of the optimization problem increases, since $Var(Y|\mat{Z})=V(\mu)\phi=\mu \exp\{\mat{g}^t(\mat{u})\mat{\gamma}\}$.

\medskip
Received September 2015 / accepted month??  year??.

\newpage
\appendix
\section*{Appendix A - Iterative process for JMMD \\ {\normalsize Adapted from \cite{PintoPonce}.}} 
\label{Appendix_A}

\noindent \textit{Mean Model} \\

Let $y_1,\cdots,y_n$ be independent observations, resulting from $n$ independent random variables $Y_1,\cdots,Y_n$;  $t_1,\cdots,t_p$ are the $p$ explanatory variables that affect the mean model and $\beta_1,\cdots,\beta_p$ are the unknown parameters of the model. Consider that $\mat{\mu}^t=(\mu_1,\cdots,\mu_n)$, $\mat{\phi}^t=(\phi_1,\cdots,\phi_n)$ and suppose that $E(Y_i|\mat{Z}_i)=\mu_i$ and $Var(Y_i|\mat{Z}_i)=\phi_iV(\mu_i)$ are known.
We start putting $k=1$, $\mat{\beta}^{t}_{0}=(0,\cdots,0)$, $\mat{\mu}^{t}_{0}=(y_1,\cdots,y_n)$ and $\mat{\phi}^{t}_{0}=(1,\cdots,1)$. Now, by using the iterative weighted least squares method, we obtain the $p\times1$ vector $\mat{\beta}_{(j)}=(\mat{T}^{t}\mat{W}_{(j-1)}\mat{T})^{-1}\mat{T}^{t}\mat{W}_{(j-1)}\mat{r}_{(j-1)}$, where the matrix $n\times p$, $\mat{T}$, is the design matrix for the mean model; the matrix $n\times n$, $\mat{W}_{(j-1)}=Diag(w_{(j-1)1},\cdots,$ $w_{(j-1)n})$, is the weight matrix for the GLM, with $Diag(.)$ representing a diagonal matrix with the elements $w_{(j-1)i}=\left(\frac{\partial \mu_{(j-1)i}}{\partial \eta_{(j-1)i}}\right)^2$ $\frac{1}{V(\mu_{(j-1)i})}$ in the diagonal, and $\mat{r}^{t}_{(j-1)}=(r_{(j-1)1},\cdots,r_{(j-1)n})$ is a $n\times 1$ vector, with $r_{(j-1)i}=\eta_{(j-1)i}+\frac{\partial \eta_{(j-1)i}}{\partial \mu_{(j-1)i}}(y_{i}-\mu_{(j-1)i})$ for $i=1,\cdots,n$ and $j=1,2,\cdots$.
In each iteration $j$ $(j=1,2,\cdots)$ a new $\mat{\beta}_{(j)}$ is obtained and the process continues until a convergence criterion is fulfilled. A possible convergence criterion could be $\|\mat{\beta}_{(j)}-\mat{\beta}_{(j-1)}\|^{2} < \delta$, where $\| \, \|$ represents the norm of a vector and $\delta \in \mathbb{R}$.
After the convergence is reached, do $\mat{W}_{k}=\mat{W}_{(j-1)}$, store the last $\mat{\beta}_{(j)}$ as $\mat{\beta}_{k}$ and use it to compute the vector $n\times 1$, $\mat{\mu}_{k}$, i.e., $\mat{\mu}_{k}=\varphi^{-1}(\mat{T}\mat{\beta}_{k})$, here $\varphi$ is an invertible known function. With the estimated value $\mat{\mu}_k$, compute the vector $n\times 1$, $\mat{d}^{*}_{k}$, whose elements are $d^{*}_{ki}=\frac{d_{ki}} {1-h_{ki}}$, where $d_{ki}=\int^{y_{i}}_{\mu_{ki}} \frac{y_{i}-\ell}{V(\ell)}d\ell$ and $h_{ki}$ is the $i$th element of the diagonal of $\mat{H}_{k}=\mat{W}^{\frac{1}{2}}_{k}\mat{T}\left(\mat{T}^{t}\mat{W}_{k}\mat{T}\right)^{-1} \mat{T}\mat{W}^{\frac{1}{2}}_{k}$. Now, knowing the $\mat{d}^{*}_{k}$ vector, we can adjust the dispersion model considering the weights $\frac{1-h_{ki}}{2}$ for each value $d^{*}_{ki}$, for $i=1,\ldots,n$. \\

\noindent \textit{Dispersion model} \\

Given $\mat{d}^{*t}_{k}=(d^{*}_{k1},\cdots,d^{*}_{kn})$ and the weights $(1-h_{k1},\cdots,1-h_{kn})$, let $u_{1},\cdots,u_{q}$ be the $q$ explanatory variables that affect the dispersion model, $\mat{\gamma}^{t}_{(0)}=(0,\cdots,0)$, $\mat{\phi}^{t}_{0}=(d^{*}_{k1},\cdots,d^{*}_{kn})$ and $\gamma_{1},\cdots,\gamma_{q}$ the unknown parameters of the model. Considering a Gamma distribution for the dispersion model and using the iterative weighted least squares method, we obtain the $q\times 1$ vector, $\mat{\gamma}_{(j)}=\left(\mat{U}^{t}\mat{V}_{(j-1)}\mat{U}\right)^{-1}\mat{U}^{t}\mat{V}_{(j-1)}\mat{s}_{(j-1)}$, where $\mat{U}$ is the $n\times q$ design matrix for the dispersion model; $\mat{V}_{(j-1)}=Diag(v_{(j-1)1},\cdots,v_{(j-1)n})$ is the $n\times n$ matrix of weights for the GLM, with $v_{(j-1)i}=\left(\frac{\partial \phi_{(j-1)i}}{\partial \xi_{(j-1)i}} \right)^{2} \frac{1}{2\phi^{2}_{(j-1)i}}(1-h_{(j-1)i})$ are the elements in the diagonal; and $\mat{s}^{t}_{(j-1)}=(s_{(j-1)1},\cdots,s_{(j-1)n})$ is an $n\times 1$ vector with $s_{(j-1)i}=\xi_{(j-1)i} + \frac{\partial \xi_{(j-1)i}}{\partial \phi_{(j-1)i}} (d^{*}_{(j-1)i}-\phi_{(j-1)i})$ for $i=1,\cdots,n$ and $j=1,\cdots$.
In the same way as it was done for the mean model, in each iteration $j$ $(j=1,2,\cdots)$ a new $\mat{\gamma}_{(j)}$ is obtained and the process continues until a convergence criterion is fulfilled. After the convergence is reached, store the last $\mat{\gamma}_{(j)}$ as $\mat{\gamma}_{k}$ and use it to compute the $n\times 1$ vector $\mat{\phi}_{k}$, that is, $\mat{\phi}_{k}=\psi^{-1}(\mat{U}\mat{\gamma}_{k})$, where $\psi$ is an invertible known function. For the dispersion model, $\psi$ is generally taken as the logarithmic function, so $\phi_k=\exp(\mat{U}\mat{\gamma}_k)$. Now, with the estimated vector $\mat{\phi}^{t}_{k}=(\phi_{k1},\cdots,\phi_{kn})$, return to the mean model and again use the iterative weighted least squares, but now with the new weights $\frac{1}{\phi_{ki}}$. Thus, in the mean model, for each $j-1$, the elements of diagonal matrix will be given by $w_{(j-1)i}=\left(\frac{\partial \mu_{(j-1)i}}{\partial \eta_{(j-1)i}}\right)^2$ $\frac{1}{\phi_{ki} V(\mu_{(j-1)i})}$. The fitting alternates between the mean and dispersion models until convergence is achieved. For example, a convergence criterion could be

\begin {equation} \label{eq16}
\frac{\left | -2Q_{Ak}^{+}-(-2Q_{Ak-1}^{+}) \right |}{\left | -2Q_{Ak}^{+}\right |}< \epsilon, \;\;\; k=1,2,\ldots
\end{equation}

\noindent where $-2Q_{Ak}^{+}$ is the extended quasi-deviance adjusted, obtained in the $k$th cycle, $\epsilon \in \mathbb{R}$ and $|\;|$ represents the absolute value operator. At the beginning of the process, consider $-2Q_{A0}^{+}=0$.

Note that in the $k$th cycle we must check if the mean and dispersion models are well adjusted and if some parameter in the mean or dispersion model should be excluded of the joint model. For more details about model checking see \cite{McCullaghNelder} and \cite{LeeNelderPawitan}.

\section*{Appendix A - Algorithm for JMMD \\ {\normalsize Adapted from \cite{PintoPonce}.}}
\label{Appendix_B}

Consider the assumptions and definitions given in Appendix A. The iterative method for the JMMD can be resumed in the following algorithm. \\

\noindent Start\\
Set $k=1$\\
Do $\mat{\beta}^{t}_{0}=(0,\cdots,0)$, $\mat{\mu}^{t}_{0}=(y_{1},\cdots,y_{n})$ and $\mat{\phi}^{t}_{0}=(1,\cdots,1)$\\
While the convergence is not achieved (joint model)\\
\vspace{0.1cm}
\hspace{1.0cm} Set $j=1$\\
\vspace{0.1cm}
\hspace{1.0cm} While the convergence is not achieved (mean model)\\
\vspace{0.1cm}
\hspace{1.5cm} Compute $\mat{\beta}_{(j)}=\left(\mat{T}^{t}\mat{W}_{(j-1)}\mat{T}\right)^{-1}\mat{T}^{t}\mat{W}_{(j-1)}\mat{r}_{(j-1)}$\\
\vspace{0.1cm}
\hspace{1.5cm} If $\|\mat{\beta}_{(j)}-\mat{\beta}_{(j-1)}\|^{2} < \delta$, stop (achieved convergence)\\
\vspace{0.1cm}
\hspace{2.5cm} Do \; $\mat{\beta}_{k}=\mat{\beta}_{(j)}$; \;
                $\mat{W}_{k}=\mat{W}_{(j-1)}$; \;
                $\mat{\mu}_{k}=\varphi^{-1}(\mat{T}\mat{\beta}_{k})$\\
\vspace{0.1cm}
\hspace{1.5cm} Else\\
\vspace{0.1cm}
\hspace{2.5cm} Do $j=j+1$\\
\vspace{0.1cm}
\hspace{1.5cm} End if\\
\vspace{0.1cm}
\hspace{1.0cm} End while\\
\vspace{0.1cm}
\hspace{1.0cm} Compute $\mat{d}^{*}_{k}$\\
\vspace{0.1cm}
\hspace{1.0cm} Do $\mat{\gamma}^{t}_{0}=(0,\cdots,0)$, $\mat{\phi}^{t}_{0}=(d^{*}_{k1},\cdots,d^{*}_{kn})$\\
\vspace{0.1cm}
\hspace{1.0cm} Set $j=1$\\
\vspace{0.1cm}
\hspace{1.0cm} While the convergence is not achieved (dispersion model)\\
\vspace{0.1cm}
\hspace{1.5cm} Compute $\mat{\gamma}_{(j)}=\left(\mat{U}^{t}\mat{V}_{(j-1)}\mat{U}\right)^{-1}\mat{U}^{t}\mat{V}_{(j-1)}\mat{s}_{(j-1)}$\\
\vspace{0.1cm}
\hspace{1.5cm} If $\|\mat{\gamma}_{(j)}-\mat{\gamma}_{(j-1)}\|^{2} < \delta$, stop (achieved convergence)\\
\vspace{0.1cm}
\hspace{2.5cm} Do \; $\mat{\gamma}_{k}=\mat{\gamma}_{(j)}$; \;
                $\mat{V}_{k}=\mat{V}_{(j-1)}$; \;
                $\mat{\phi}_{k}=\psi^{-1}(\mat{U}\mat{\gamma}_{k})$\\
\vspace{0.1cm}
\hspace{1.5cm} Else\\
\vspace{0.1cm}
\hspace{2.5cm} Do $j=j+1$\\
\vspace{0.1cm}
\hspace{1.5cm} End if\\
\vspace{0.1cm}
\hspace{1.0cm} End while\\
\vspace{0.1cm}
\hspace{1.0cm} Do $-2Q_{A0}^{+}=0$\\
\vspace{0.1cm}
\hspace{1.0cm} If $\frac{| -2Q_{Ak}^{+}-(-2Q_{Ak-1}^{+})|}{|-2Q_{Ak}^{+}|} < \epsilon$, stop (achieved convergence)\\
\vspace{0.1cm}
\hspace{1.0cm} Else\\
\vspace{0.1cm}
\hspace{2.5cm} Do $k=k+1$\\
\vspace{0.1cm}
\hspace{1.0cm} End if\\
\vspace{0.1cm}
End while\\
\vspace{0.1cm}
Do $\mat{\beta}=\mat{\beta}_{k}$; \; $\mat{\gamma}=\mat{\gamma}_{k}$\\
\vspace{0.1cm}
End

\end{document}